\documentclass[aps,prl,twocolumn,tightenlines,showpacs,aps,amsmath,amssymb,nofootinbib]{revtex4-1}

\usepackage{latexsym}
\usepackage{graphicx}
\usepackage{subfigure}
\usepackage{cases}
\usepackage{hyperref}
\usepackage{amssymb}
\usepackage{bm}
\usepackage{natbib}
\usepackage{amsmath}
\usepackage{slashed}
\usepackage{braket}
\usepackage{color}  
\usepackage{youngtab}

\newcommand{\be}{\begin{equation}}
\newcommand{\ee}{\end{equation}}
\newcommand{\ba}{\begin{eqnarray}}
\newcommand{\ea}{\end{eqnarray}}
\def\bal{\begin{align}}
\def\eal{\end{align}}
\def\bald{\begin{aligned}}
\def\eald{\end{aligned}}

\begin{document}

\title{A possible phase for adjoint QCD}

\date{\today}

\author{Erich Poppitz}
\email[]{poppitz@physics.utoronto.ca}
\affiliation{Department of Physics, University of Toronto, Toronto, ON M5S 1A7, Canada}
\author{Thomas A. Ryttov}
\email[]{ryttov@cp3.sdu.dk}
\affiliation{CP$^3$-Origins, University of Southern Denmark, Campusvej 55, 5230 Odense M, Denmark}


\begin{abstract}
We discuss an exotic phase that adjoint QCD possibly exhibits in the deep infrared (IR). It is a confining phase, with a light spectrum consisting of massless composite fermions. The  discrete chiral symmetry is broken, with unbroken continuous chiral symmetry. We argue that it may give a description of the IR of adjoint QCD with three massless Weyl flavors and that 
it passes all consistency checks known to us.
\end{abstract}


\maketitle

{\flushleft\textbf{Introduction:}}  In this note, we discuss an exotic phase that adjoint QCD possibly exhibits in the deep infrared (IR). As explained in the abstract, this phase has an IR behavior that is quite simple to describe, yet has not been discussed in the literature. It is worth filling this gap, especially in light of the recent discussions of  several conjectured exotic IR phases of  two-flavor  two-color adjoint QCD \cite{Anber:2018tcj,
Cordova:2018acb,Bi:2018xvr,Wan:2018djl}.

Before we continue with a detailed discussion of the proposed phase, we note that studies of adjoint QCD have been motivated by  a wide range of interests:  possible applications in beyond the standard model physics, field-theoretic studies of confinement,    phenomena emerging in the limit of large number of colors, and even novel many-body phases (an incomplete list of references is \cite{Bi:2018xvr,Wan:2018djl,Sannino:2004qp,Catterall:2007yx,Unsal:2007vu,Unsal:2007jx,Cherman:2018mya}).

{\flushleft\textbf{Theory and symmetries:}}  Adjoint QCD is an $SU(n_c)$ gauge theory with a set of $n_f$ Weyl fermions in the adjoint representation of the gauge group. We  denote the gauge fields as $A_{\mu} = A_{\mu}^a T^a$, where $T^a$, $a=1,\ldots, n_c^2-1$, are the generators of the gauge group while $F_{\mu\nu} = F^a_{\mu\nu}  T^a  $ is the associated field strength tensor. We refer to the gauge fields as gluons and we  borrow language from supersymmetry and refer to the adjoint Weyl fermions $\lambda_{\alpha}^i = \lambda^{a,i}_{\alpha} T^a$, $i=1,\ldots n_f$, as gluinos. Lastly, $\mu,\nu,\ldots $ are Lorentz indices while $\alpha,\beta,\ldots$ are SL$(2,\mathbb{C})$ spinor indices. 

The theory is asymptotically free for a range of number of fermion flavors, $0 \leq n_f  \leq 5$, where we want to study the deep IR physics of the theory. Starting from the bottom up, the $n_f=0$ theory is pure Yang-Mills theory which  confines and generates a mass gap.  The $n_f=1$ theory is pure $\mathcal{N}=1$ supersymmetric Yang-Mills theory, which also confines and breaks its discrete chiral symmetry.

On the other hand if we instead start from the top both the $n_f=5$ and $n_f=4$ theories are believed to flow to a non-trivial IR fixed point \cite{Caswell:1974gg,Banks:1981nn}. This conclusion is in very good agreement with the majority of lattice simulations, see for instance \cite{Catterall:2007yx,DelDebbio:2010hx,Hietanen:2009az,Bergner:2016hip}, as well as recent analytical computations \cite{Ryttov:2016hdp,Ryttov:2017kmx,Ryttov:2017lkz}. As a consistency check on this picture one can compute the anomalous dimension $\gamma$ of the fermion bilinear in a scheme independent and unambiguous manner order by order in perturbation theory and then perform a Pad\'e extrapolation to get a feel for the all orders exact result. In the case $n_c=2$ one finds $\gamma\sim 0.128 $ for $n_f=5$ and $\gamma \sim 0.548$ for $n_f=4$ \cite{Ryttov:2017lkz}. Both values are reasonably small and well within the unitarity bound $\gamma<2$ so one expects that the phase with a non-trivial IR fixed point is indeed realized. According to standard lore many of these features of adjoint QCD are expected to have only little sensitivity to the number of colors $n_c$. 

The question remains, however, what happens to the $n_f=3$ theory in the middle range, for which much less is known. Only a few lattice simulations exist \cite{Bergner:2017gzw} and it is fair to say that our knowledge of the IR phase of the theory is still very limited. For instance if one assumes that the theory is at a non-trivial IR fixed point then the Pad\'e extrapolated all orders result is $\gamma \sim 1.59$ for the anomalous dimension of the fermion bilinear \cite{Ryttov:2017lkz}. This value is considerably large (and close to the unitarity bound) and we therefore expect the theory to display an IR phase different from the $n_f=4$ and $n_f=5$ theories. It is for this $n_f=3$ theory that we shall propose an exotic candidate phase. For now however we shall stick to keeping $n_f$ arbitrary although we will eventually settle at $n_f=3$. We shall have  little to say about the $n_f=2$ theory; a nice summary of the various phases proposed recently is in \cite{Wan:2018djl}.

Adjoint QCD has a number of global symmetries. At the classical level there is a 0-form global symmetry $G^{(0)} = SU_f(n_f) \times U(1)$, where the $U(1)$ is anomalous and is broken to a discrete subgroup so that the theory has a 0-form global symmetry $G^{(0)} = SU_f(n_f) \times \mathbb{Z}_{2n_fn_c}$; identifications of the center of $SU_f(n_f)$ with a subgroup of $\mathbb{Z}_{2n_fn_c}$ are not shown but will be discussed later. The 0-form symmetry acts on local degrees of freedom (fields) in the theory.

Since the fermions are in the adjoint representation, the theory also has a 1-form discrete center symmetry $G^{(1)}=\mathbb{Z}^{(1)}_{n_c}$. The center symmetry acts on line operators and is related to confinement---if the theory confines fundamental charges, the center symmetry is unbroken.

{\flushleft\textbf{Summary of the proposed IR phase:}} We want to suggest that in the deep IR the theory $i.)$ confines, or has an unbroken center symmetry, $ii.)$ has massless composite fermions, and $iii.)$ has broken discrete chiral symmetry but manifest continuous chiral symmetry. The first point is an assumption,  the second is taken care of by finding a set of gauge invariant operators creating composite massless excitations which satisfy 't Hooft anomaly matching of all the 0-form global symmetries, and the last point ensures the 1-form symmetry anomaly matching. Next, we discuss the proposal in   more detail.

{\flushleft\textbf{Massless spectrum and 0-form anomaly matching for $\mathbf{n}_\mathbf{c}$ $\mathbf{=}$ $\mathbf{2}$:}} 
 We begin with our proposal for a massless IR spectrum. It is very simple and in a certain sense is a (gauge invariant) ``copy" of the ultraviolet fermionic degrees of freedom. This ensures that all 0-form 't Hooft anomalies are automatically satisfied. 

To be specific, consider the $n_c=2$ case while for the moment still keeping $n_f$ arbitrary. We propose that the infrared is described by the following set of massless composite fermions
\begin{eqnarray}\label{ops1}
\left( {\cal O}_1 \right)_{\alpha}^{i} &=& \text{Tr} \left[ F_{\mu\nu} (\sigma^{\mu\nu})_{\alpha}^{\phantom{\alpha} \beta}  \lambda_{\beta}^i \right]    ~,\nonumber \\
\left(  {\cal O}_2 \right)_{\alpha}^{i} &=&  \text{Tr} \left[  F_{\mu\nu}   F^{\nu}_{\phantom{\nu}\rho}(\sigma^{\rho\mu})_{\alpha}^{\phantom{\alpha} \beta} \lambda_{\beta}^{i}  \right]  ~, \\
\left(  {\cal O}_3 \right)_{\alpha}^{i} &=&  \text{Tr} \left[ F_{\mu\nu} F^{\nu \rho} F_{\rho\sigma} (\sigma^{\sigma\mu})_{\alpha}^{\phantom{\alpha} \beta} \lambda_{\beta}^i  \right]_{\alpha}^i ~, \nonumber 
\end{eqnarray}
where the trace is over the adjoint color indices and the Lorentz indices are contracted all the way through in a connected manner.

As can be seen, the proposal essentially consists of the gluinos dressed with an appropriate amount of glue. Since they each contain only a single $\lambda^i$ they are all fundamentals under $SU_f(n_f)$ flavor and carry charge $+1$ under discrete chiral $\mathbb{Z}_{4n_f}$ (recall that $n_c=2$). In Table \ref{SymmetryAdjointQCD} we have summarized the symmetries and charges of both the ultraviolet (UV) and IR fermions. 

Since  our proposal for the IR spectrum is  a   gauge invariant copy of the UV massless fermions, all 't Hooft anomalies of the 0-form global  symmetry $G^{(0)} = SU_f(n_f) \times \mathbb{Z}_{4n_f }$ are matched between the UV and IR, including the relevant mixed anomalies with gravity. For completeness, let us enumerate, for $n_f>2$, the non-vanishing anomalies matched by the massless fermions---$\lambda$ in the UV,  and ${\cal O}_{1,2,3}$ in the IR---
$ \left[ SU_f(n_f) \right]^3$, $
 \left[ \mathbb{Z}_{4n_f} \right]^3$, $
\mathbb{Z}_{4n_f} \left[ SU_f(n_f) \right]^2$, $
 \mathbb{Z}_{4n_f} \left[ G \right]^2$, where $G$ denotes gravity (for $n_f=2$, the $\left[SU_f(n_f)\right]^3$ anomaly should be replaced by the Witten anomaly). 
\begin{table}
\begin{center}
\begin{tabular}{|c|c|c|c||c|} 
\hline\hline
   & $SU(2)$  &  $SU_f(n_f)$ & $\mathbb{Z}_{4n_f}$ & $\mathbb{Z}_{2n_f}$       \\ 
\hline
$\lambda$  & adj  & $\tiny{\yng(1)} $ & $1$ & $1$    \\
\hline 
${\cal O}_1$ & $1$ & $\tiny{\yng(1)}$ & $1$ & $1$  \\
${\cal O}_2$ & $1$ & $\tiny{\yng(1)}$ & $1$ & $1$  \\
${\cal O}_3$ & $1$ & $\tiny{\yng(1)}$ & $1$ & $1$  \\
\hline
\hline
\end{tabular}
\end{center}
\caption{0-form symmetries of adjoint QCD for $n_c=2$.}
\label{SymmetryAdjointQCD}
\end{table}

 The ability to make a ``gauge invariant copy" of the gluinos is due to their adjoint nature.\footnote{Similar gauge invariants appear in the descriptions of supersymmetric theories with adjoint matter fields, as in \cite{Cachazo:2002ry}.}
The assumed accidental degeneracy and independence of the states created by ${\cal O}_{1,2,3}$, is an admitted weakness of our proposal. Our main defense  is its consistency with anomalies and  simplicity. 

To stress the latter point, we first recall  that for  $n_f=2$ there is a  more minimal, compared to (\ref{ops1}), solution of the 0-form anomaly matching conditions in terms of a single massless composite $SU_f(2)$-doublet, schematically, a  Tr($\lambda  \lambda \lambda$) ``baryon," see  \cite{Anber:2018tcj}, essentially because the $SU_f(2)$ Witten anomaly is $\mathbb{Z}_2$-valued and hence less restrictive.

For $n_f=3$, however, the anomaly constraints are significantly tighter. Using the cubic Dynkin indices for $SU_f(3)$, see \cite{Ramond:2010zz}, one can  construct other solutions. A simple example is a spectrum of massless Weyl fermions consisting of a three-index symmetric tensor of $SU_f(3)$ (contained in  $\lambda \lambda F_{\mu\nu} \sigma^{\mu\nu}\lambda $, a ``baryon"  in the $\mathbf{10}$ of $SU_f(3)$ with cubic Dynkin index $27$ and $\mathbb{Z}_{12}$ charge 3) plus 24 $SU_f(3)$ antifundamentals (or $\mathbf{3^*}$, schematically, of the form Tr($\sigma^\mu \bar \lambda F^k_{...} D_\mu F^p_{...}$), with some choice of integer $k$ and $p$, of cubic Dynkin index $-1$ and $\mathbb{Z}_{12}$ charge -1). This massless spectrum  matches the $[SU_f(3)]^3$ and  the $\mathbb{Z}_{6}$ $[SU_f(3)]^2$ anomalies of the UV theory,  but fails to match the mixed $\mathbb{Z}_{6}$ $[G]^2$ anomaly.\footnote{The matching of the $[SU_f(3)]^3$ anomaly follows upon adding the cubic Dynkin indices of the massless fermions.  The anomalies involving  $\mathbb{Z}_{6}$ (the unbroken subgroup of the $\mathbb{Z}_{12}$ chiral symmetry, see next section) should be matched modulo $6$, but it is easy to check that they match modulo $12$ as well, as for (\ref{ops1}). The UV contribution to the  $\mathbb{Z}_{6}$ $[SU_f(3)]^2$  anomaly from the UV Weyl fermions   is $3$ (counting the number of zero modes in an $SU_f(3)$ instanton \cite{Csaki:1997aw}), while in the IR
the $\mathbf{10}_3$ contributes $3 \times 15$, where $15$ is the quadratic Dynkin index of the $\mathbf{10}$ of $SU_f(3)$, and the 24 $\mathbf{3^*}_{-1}$ fermions contribute $-24$, for a total of $45-24=21 = 3$ (mod$6$). For the $\mathbb{Z}_{6}$ $[G]^2$ anomaly, the UV contribution is $9$ (counting the number of zero modes in a  gravitational instanton background), while the massless IR spectrum contributes $3 \times 10$ from the $\mathbf{10}_3$ and $- 24$  from the $\mathbf{3}_{-1}$, for a total IR contribution of  $6$ $\ne 9 ({\rm mod} 6)$. The flavor-singlet $\mathbf{1}_3$ baryon gives an additional factor of $3$, bringing the IR contribution to $9$, thus matching the UV anomaly. Finally,  the $\mathbb{Z}_6^3$ anomaly is also matched: in the UV it equals $9$, while in the IR it is $3^3\times 10 - 24 + 3^3 = 6\times 45 + 3 = 9 \;({\rm mod}6)$. }  Matching of the latter anomaly as well as of the $[\mathbb{Z}_{6}]^3$ is achieved by adding an $SU_f(3)$-singlet baryon (a $\mathbf{1}$ contained in  $\lambda \lambda \lambda$), a Weyl fermion of charge $3$. We can summarize this solution of 0-form anomaly matching as having a massless  spectrum of Weyl fermions transforming as
\begin{equation}
\label{eq3}
\mathbf{10}_3 + 24 \times \mathbf{3^*}_{-1} + \mathbf{1}_{3}~
\end{equation}
 under $SU_f(3)_{\mathbb{Z}_6}$.
As  the values of the cubic Dynkin indices increase  quite rapidly with the dimension of $SU_f(3)$ representations, it follows  that  solutions other than (\ref{ops1}) and (\ref{eq3}), which no doubt exist,    would necessarily be more baroque. Comparing  (\ref{ops1}) with (\ref{eq3}) also reveals that the former is more minimal.\footnote{A comparison based on the $a$-theorem \cite{Komargodski:2011vj} shows that both (\ref{ops1}) and (\ref{eq3}) are consistent with it, with  smaller $a_{IR}$   for (\ref{ops1}). To verify this, one only needs to note that the contributions to $a$ from massless vectors and Weyl fermions obey $a_{vector}/a_{Weyl} = 124/11$.}  The spectrum (\ref{ops1}) also has the advantage that it is easily generalized to arbitrary $n_c$.

Further support for our proposed spectrum may be derived from the fact  that the assumed global symmetry realization (with unbroken continuous chiral symmetry and broken discrete chiral symmetry)  is not a complete fantasy  but is actually the one observed in a theoretically controlled study of adjoint QCD. Analytic control over the IR of the theory is achieved upon compactification to  $\mathbb{R}^{1,2} \times \mathbb{S}^1$, with periodic boundary conditions for the fermions on the $\mathbb{S}^1$ and with the circle size taken small compared to the strong coupling scale of the theory.
In the limit    of small $\mathbb{S}^1$, adjoint QCD  can be solved semiclassically with the result  that the symmetries are realized exactly as in our proposal   \cite{Unsal:2007vu,Unsal:2007jx}. 
Thus, our proposal is consistent with a flow,  upon compactification to small $\mathbb{S}^1$,  with no change of symmetry realization. Our conjectured massless spectrum, however, changes upon reducing the   $\mathbb{S}^1$ size---two of the three massless $SU(n_f)$ fundamentals of  (\ref{ops1}) obtain chirally symmetric masses at small $\mathbb{S}^1$ from their coupling to the center symmetric holonomy. 
  
{\flushleft\textbf{Symmetry realization and anomaly matching  involving the 1-form symmetry for $\mathbf{n}_\mathbf{c}$ $\mathbf{=}$ $\mathbf{2}$:}} It was recently realized that the gauging of 1-form symmetries leads to new 't Hooft anomaly matching conditions \cite{Gaiotto:2014kfa,Gaiotto:2017yup}. For the theory at hand, the relevant anomaly is the mixed $\mathbb{Z}_{4 n_f}$-$\mathbb{Z}_{2}^{(1)}$ discrete chiral/1-form center anomaly.

To detect this anomaly, one  introduces a two-form center symmetry gauge field background, which has the effect of changing the  topological charge quantization from integer to half-integer \cite{Gaiotto:2017yup}. Equivalently, gauging the  center symmetry of an $SU(2)$ theory amounts to introducing 't Hooft flux backgrounds, whose topological charge is  half-integer \cite{tHooft:1979rtg,vanBaal:1982ag}. The half-integer quantization of topological charge now  implies that the partition function of the fermions acquires a $\mathbb{Z}_2$ phase under a discrete chiral $\mathbb{Z}_{4n_f}$ transformation. This phase represents the mixed 't Hooft anomaly and has to be matched between the UV and IR of the theory. Our proposed IR spectrum only involves the  free fermions created by the local operators ${\cal O}_{1,2,3}$. As these local fields are invariant under the 1-form symmetry, anomalies involving center symmetry have to be matched by some other means. 

As in \cite{Anber:2018tcj}, we propose that they are matched in the ``Goldstone" mode, i.e.~that the discrete chiral  $\mathbb{Z}_{4n_f}$ suffers a spontaneous breakdown  $\mathbb{Z}_{4n_f} \rightarrow \mathbb{Z}_{2 n_f}$, due to the $SU_f(n_f)$ invariant expectation value 
$\langle {\rm det}_{ij} (\epsilon^{\alpha\beta } \lambda^{i \; a}_\beta \lambda^{j \; a}_\alpha) \rangle$. The anomaly is matched in the IR by an IR topological quantum field theory (TQFT). This is a theory with a finite dimensional Hilbert space, a kind of ``chiral lagrangian" describing the two degenerate ground states resulting from the breaking of the discrete chiral symmetry. Such TQFTs arising from the spontaneous breaking of discrete chiral symmetries and matching the relevant mixed anomalies with 1-form symmetries have been explicitly described in  \cite{Gaiotto:2014kfa,
Anber:2018xek}.

Thus, a complete description of the proposed IR phase can be summarized as the theory of the free massless fermions ${\cal O}_{1,2,3}$ tensored with the TQFT resulting from the $\mathbb{Z}_{4n_f} \rightarrow \mathbb{Z}_{2 n_f}$ chiral symmetry breaking.

For the case of $n_f=2$  \cite{Anber:2018tcj}, after the work \cite{Cordova:2018acb} (see also \cite{Bi:2018xvr,Wan:2018djl})  it is known that 
 the TQFT arising from the $\mathbb{Z}_{4n_f} \rightarrow \mathbb{Z}_{2 n_f}$  spontaneous discrete chiral symmetry breakdown does not suffice to match all 't Hooft anomalies involving 1-form symmetries and that the $n_f=2$ theory has  't Hooft anomalies not captured by the above analysis.
  This is because the two-flavor adjoint theory is special---it  exhibits the so-called ``nonabelian spin-charge relation." This relation is nothing but the  statement that all states, or gauge invariant operators, that are in half-integer-spin $SU_f(2)$  representations are fermions, while bosons have integer $SU_f(2)$ spin.\footnote{Equivalently, the spin-charge relation is the realization that the $\mathbb{Z}_2$ center of $SU_f(2)$ is identified with fermion parity $(-)^F$.} 
  
  From the point of view of 't Hooft anomalies, the importance of the nonabelian spin-charge relation is that it allows the theory to be formulated on non-spin manifold backgrounds.  The $\mathbb{Z}_2$ obstruction to globally defining spinors on such manifolds is compensated by turning on appropriate  $\mathbb Z_2$ twists of the $SU_f(2)$ background fields (the latter are ``felt" by the fermions in the half-integer spin representation but not by the bosons in integer-spin representations, thus allowing for a consistent formulation of the full theory on such backgrounds).  It was  further  found that the UV theory exhibits  't Hooft anomalies detected only in such backgrounds. Renormalization group invariance then requires that these anomalies should be reproduced by any conjectured IR theory (which necessarily also obeys the spin-charge relation) when placed in the same background, resulting  in  a new anomaly matching condition. 

 For the $n_f=2$ theory at hand,  the 't Hooft fluxes' topological charge on non-spin manifolds  is quantized in units of $1/4$ instead of $1/2$, and the mixed $\mathbb{Z}_8$-$\mathbb{Z}_2^{(1)}$ anomaly is not saturated by the TQFT describing the $\mathbb{Z}_8 \rightarrow \mathbb{Z}_4$ breaking. This requires  additional degrees of freedom to match the anomaly.\footnote{See \cite{Cordova:2018acb} for other anomalies involving $\mathbb{Z}_2^{(1)}$ and geometry.} Such additional  IR TQFTs for the $n_f=2$ theory in the $\mathbb{Z}_8 \rightarrow \mathbb{Z}_4$ phase have been constructed \cite{Wan:2018djl}. However, their possible UV origin remains mysterious to us. This makes it hard to judge  whether any  of the  existing proposals for exotic (i.e. without continuous chiral symmetry breaking) phases enumerated in \cite{Wan:2018djl} represent the true IR behavior of the two-flavor theory. 
 
In contrast,  $n_f=3$ adjoint QCD exhibits no  spin-charge relation (the center of $SU_f(3)$ is $\mathbb{Z}_3$, not identified with $(-)^F$ fermion parity) and the associated additional anomaly constraints are absent. As already noted, our conjectured phase is for $n_f=3$.

{\flushleft\textbf{Generalizing to arbitrary $\mathbf{n}_\mathbf{c}$:}}  So far we have discussed the $n_c=2$ case but the generalization of the IR spectrum matching the 0-form anomalies (\ref{ops1}) to arbitrary finite $n_c$ is straightforward. We propose that the massless IR spectrum is composed of $n_c^2-1$ gauge invariant operators made up of gluinos dressed with an appropriate amount of glue:
\begin{eqnarray}\label{ops}
\left( {\cal O}_1 \right)_{\alpha}^{i} &=& \text{Tr} \left[ F_{\mu\nu} (\sigma^{\mu\nu})_{\alpha}^{\phantom{\alpha} \beta}  \lambda_{\beta}^i \right] ~, \nonumber  \\
&\vdots&
 \\
\left(  {\cal O}_{n_c^2-1} \right)_{\alpha}^{i} &=&  \text{Tr} \big[ \underbrace{F_{\mu\nu} \cdots F_{\rho\sigma}}_{n_c^2-1} (\sigma^{\sigma\mu})_{\alpha}^{\phantom{\alpha} \beta} \lambda_{\beta}^i  \big]_{\alpha}^i ~, \nonumber 
\end{eqnarray} 
where again the trace is over the adjoint color indices and the Lorentz indices are contracted all the way through in a connected manner (discussion of the many questions that may arise, concerning minimality, uniqueness, large-$n_c$ limit, etc., follows further below). It is clear  that, just as in the $n_c=2$ case,  these $n_c^2-1$ gauge invariant operators precisely mimic the UV fermionic degrees of freedom. They are each a fundamental of $SU_f(n_f)$ flavor and carry charge $+1$ under the discrete chiral $\mathbb{Z}_{2n_fn_c}$. Therefore again all the 't Hooft anomalies of the 0-form global symmetry are automatically matched between the UV and IR. 
We propose that these are the massless degrees of freedom adjoint QCD exhibits in the deep IR for $n_f=3$ and any (finite) $n_c$.

In order to match the mixed $\mathbb{Z}_{2 n_fn_c}$-$\mathbb{Z}_{n_c}^{(1)}$ discrete chiral/center anomaly in the IR we also again propose the formation of the $SU_f(n_f)$ flavor invariant condensate $\langle {\rm det}_{ij} (\epsilon^{\alpha\beta } \lambda^{i \; a}_\beta \lambda^{j \; a}_\alpha) \rangle$ that breaks the discrete chiral symmetry $\mathbb{Z}_{2n_fn_c} \rightarrow \mathbb{Z}_{2 n_f}$. As was the case for $n_c=2$, the massless field content of the IR theory should be supplemented with an IR TQFT describing the $n_c$ vacua due to the    $\mathbb{Z}_{2n_fn_c} \rightarrow \mathbb{Z}_{2 n_f}$ breaking and matching the $\mathbb{Z}_{2 n_fn_c}$-$\mathbb{Z}_{n_c}^{(1)}$ anomaly.

Similar to the $n_c=2$ case, the main argument in favor of our proposal is consistency of the spectrum and symmetry realization with all  anomaly matching conditions known to us. 
However, the proposed IR spectrum (\ref{ops}), especially when considered for arbitrary large $n_c$,  has interesting features and raises questions, some of which  remain unanswered. We now discuss these, focusing on  $n_f=3$:

{\flushleft{\bf{1.} {\it Non-uniqueness.}}} We do not expect that the solution (\ref{ops}) is   unique. One can construct other  massless spectra matching the anomalies using $SU_f(3)$ representations different from the $\mathbf{3}$, in a manner  similar to how we proceeded  for $n_c=2$ above to arrive at the massless fermion spectrum (\ref{eq3}). 

{\flushleft{\bf{2.} {\it Large multiplicities and large anomalous dimensions.}}} Despite the non-uniqueness, one feature of (\ref{ops}) is likely to be generic to any proposed solution of anomaly matching:  the large, ${\cal{O}}(n_c^2$),  multiplicity of identical representations, implying that the operators in  (\ref{ops}) must acquire large anomalous dimensions to become free in the IR. 

 To argue this, note that the reason we require $n_c^2-1$ copies of  $\mathbf{3}$ (of cubic Dynkin index $1$) is that the $[SU_f(3)]^3$ anomaly in the UV theory is $n_c^2 -1$. Whatever other massless representations for Weyl fermions one contemplates, the sum of their cubic Dynkin indices has to add to $n_c^2-1$. If we take the limit of  large (infinite) $n_c$, and if we keep the maximum number of boxes in the Young tableau of the $SU_f(3)$ representations of the  massless fermions finite as we take $n_c$ large, their cubic Dynkin indices will be finite, and ${\cal{O}}(n_c^2$) multiplicities will become necessary to match the $[SU_f(3)]^3$ anomaly.\footnote{One can imagine relaxing this assumption and allowing the number of boxes in the Young tableau of the $SU_f(3)$ representation to grow with $n_c$. We have not studied this in any detail and only note that the relevant composite operators will now have up to $n_c$ insertions of $\lambda^i$, even before taking into account any multiplicities needed to match the $[SU_f(3)]^3$ anomaly.} Further, in order to have ${\cal{O}}(n_c^2$) multiplicities of an identical representation, the operators that create the different massless states will have to invoke insertions of $F^k$, $k=1,..., {\cal{O}}(n_c^2)$ ($F$ stands for field strength tensors or derivatives), as in (\ref{ops}). While the above discussion does not constitute a proof, it suggests that operators of increasingly higher classical dimension (in (\ref{ops}), up to  ${\cal{O}}(n_c^2)$) must acquire large anomalous dimensions to become free in the IR.

This  may appear unusual, but we note that
  similar features have been seen in  theories studied earlier. Consider  
Seiberg's duality of $SU(n_c)$ ${\cal{N}}=1$ supersymmetric QCD with $n_f = n_c+1$  fundamental flavors \cite{Seiberg:1994pq}. The IR phase of these theories has a somewhat similar flavor to our proposed IR phase. First, these are also phases without  continuous global chiral symmetry breaking. The massless composite supermultiplets, the mesons $M_i^j$ and baryons $b^i$ and $\tilde{b}_j$ ($i,j=1,\ldots, n_c+1$), saturate 't Hooft anomaly matching at the origin of moduli space. Second, the number of massless composites is ${\cal{O}}(n_c^2)$, similar to  our proposed phase. The meson and baryon fields are free in the deep IR and there is a large emergent global symmetry with rank  $\sim n_c^2$,   somewhat  similar to our proposal.\footnote{The emergent global symmetry is broken by the interactions (irrelevant in the IR),  for example those given by the superpotential $W={\rm det} M + b  M  \tilde{b}$,  down to the flavor symmetry group. We also expect this to be the case for the (unknown) interactions between the composite fermions (\ref{ops}).} Lastly, the composite operators $M, b, \tilde{b}$ develop  large anomalous dimensions, as the meson and baryon superfields have classical dimensions $2$ and $n_c$, respectively, while both become free fields in the IR, as should our operators (\ref{ops}). 

Clearly, the similarity is at most suggestive. In particular, we lack the ``power of supersymmetry" and    the impressive checks of Seiberg's duality showing consistency with renormalization flows upon symmetry-breaking (flat-direction) or mass deformations  cannot be performed with any degree of confidence, not the least because phase transitions upon deformations cannot be ruled out, but are likely  to be present, in non-supersymmetric theories.

{\flushleft{\bf{3.} {\it Large-$n_c$ limit.}}} The operators in (\ref{ops})---and as we argued above, likely in any other solution of anomaly matching via composite fermions---are schematically of the form Tr$(F^p ...)$, where $p$ can be as large as $\sim n_c^2$.  Usual large-$n_c$ counting rules hold for operators where $p$ is kept fixed as $n_c \rightarrow \infty$. Notable exceptions are baryon operators  in large-$n_c$ QCD, which are understood  
 \cite{Witten:1979kh}, albeit the physical intuition involved does not apply to the case at hand. Of closer relevance is the case of  supersymmetric QCD with $n_f = n_c+1$, discussed above, where the massless baryons $b, \tilde{b}$   involve the product of $n_c$ quark superfields; however, we are not aware of a study of the large-$n_c$ limit of Seiberg's theory which discusses  baryons.\footnote{The only study of large-$n_c$ counting rules in Seiberg duality known to us \cite{Schmaltz:1998bg} considered only the large-$n_c$   mesons.} 
 
 Thus,  regarding the fate of our proposal in the large-$n_c$ limit, we can only note that either the counting is modified to accommodate the operators in (\ref{ops}) (as it probably is for the baryons in supersymmetric QCD) or there is an inconsistency (or perhaps a large-$n_c$ phase transition)  with our proposed symmetry realization and spectrum in the infinite $n_c$ limit. This is a question worthy of future study.
 
 {\flushleft{\bf{4.} {\it ``Daughter-daughter" planar equivalence.}}}  Related to the above discussion, it is interesting to juxtapose our proposal with the   large-$n_c$ planar equivalence between two theories, denoted as A and B  below \cite{Armoni:2003fb,Armoni:2003gp}. Theory A is the one we study here---adjoint QCD with $n_f$ Weyl flavors  and an $SU(n_c)$ gauge group. Theory B has also an $SU(n_c)$ gauge group, but with $n_f$ Dirac flavors in the two-index antisymmetric (or symmetric) representation. It has been argued, assuming unbroken charge conjugation on $\mathbb{R}^4$ \cite{Unsal:2006pj}, that there is a planar equivalence between the bosonic sectors of the  theories $A$ and $B$.  
 This already implies that the spectrum of fermionic operators ${\cal O}_i$, $i=1,\ldots,n_c^2-1$ of theory $A$ cannot be probed using planar equivalence. 
Despite this, the bosonic equivalence alone suggests that the continuous\footnote{The matching of discrete symmetry is more subtle. To make the case note that, at finite $n_c$, theory A has a $\mathbb{Z}_{n_c}^{(1)}$ symmetry while theory B only has a $\mathbb{Z}_{2}^{(1)}$ for even $n_c$; notice however that center symmetries may emerge at infinite $n_c$  \cite{Armoni:2007kd}. } global symmetry realization of the two theories is the same in the infinite $n_c$ limit.  

One can try to use this equivalence to argue as follows. Let us  assume  that theory B behaves similarly at $n_c\rightarrow \infty$ and $n_c=3$. Next, we use  the equivalence of theory B at $n_c=3$ to three-flavor massless QCD with fundamental Dirac fermions, which is believed to break chiral symmetry.\footnote{Anomaly matching for $SU(3)$ QCD with $n_f = 3 k$ was used to argue  that  chiral symmetry is broken \cite{Preskill:1981sr}. While an IR CFT was not discussed there as  a possibility, the absence of solutions to anomaly matching argued in \cite{Preskill:1981sr} suffices to  exclude a confining phase without chiral symmetry breaking for $n_f =3k$.} Thus, by our assumption, we are led to conclude that at $n_c \rightarrow \infty$ the continuous chiral symmetry in theory B should be broken. Then the daughter-daughter planar equivalence would require that theory A also breaks its continuous flavor symmetry in the planar limit, in effect implying that our proposal does not hold at $n_c \rightarrow \infty$. 
While  this  conclusion might  be true, the main assumption, that theory B behaves similar at $n_c=3$  to $n_c \rightarrow \infty$, has potential pitfalls, as there may be a phase transition on the way to infinite $n_c$,  examples of such behavior can be constructed.\footnote{A strong case to consider is the $n_f=6$ theory. At $n_c=3$ theory B is $n_f=6$ flavor QCD which is believed to break continuous chiral symmetry since there does not exist a gauge invariant set of composite operators which saturates all the 0-form global 't Hooft anomalies \cite{Preskill:1981sr}. However as $n_c \rightarrow \infty$ the theory looses asymptotic freedom and is instead IR free (it has also moved trough the conformal window on the way). So there are two phase transitions on the way to infinite $n_c$ in theory B. Theory A is similarly non-asymptotically free and hence IR free for all $n_c$. If we lower $n_f$ and consider $n_f=5$, a single phase transition exists. At $n_c = 3$ theory B is QCD with five flavors which is also believed to break chiral symmetry through the formation of a bilinear condensate. On the other hand at $n_c=2, 3,...$ theory A with $n_f=5$ is believed to be at a non-trivial IR fixed point without a bilinear condensate and this is expected to hold for any $n_c$.  So as we depart from $n_c=3$ in theory B ($n_f=5$) we enter a new phase (the conformal window) at some finite $n_c$ before we arrive at infinite $n_c$. The details of what specific value of $n_c$ we enter the conformal window is inessential;    it happens somewhere along the way in theory B as we move $n_c$ from three towards infinity. The same picture should also be valid for the $n_f=4$ case since there theory A is also believed to be at a non-trivial IR fixed point without a bilinear condensate and that feature is also expected to hold at any $n_c$.  }

\smallskip
 
\begin{table*}[t]
\begin{center}
\begin{tabular}{|c|c|c|c|c|} 
\hline\hline
  $n_f$ & IR Phase  &  Intact c$\chi$ sym.  &  Intact d$\chi$ sym. &  Intact center sym.      \\ 
\hline
$\geq 6$  & Free  & Yes & Yes & No    \\
$5$ & Fixed point & Yes & Yes & No  \\
$4$ & Fixed point & Yes & Yes & No  \\
$3$ & Confinement, massless composite fermions & Yes & No & Yes  \\
$2$ & Confinement &  No & No &  Yes  \\
$1$ & {\cal N} =1 SYM & --- & No & Yes  \\
$0$ & Pure YM & --- & --- & Yes  \\
\hline
\hline
\end{tabular}
\end{center}
\caption{The IR phases of adjoint QCD with $n_f$ Weyl flavors. }
\label{IRphases}
\end{table*}

{\flushleft\textbf{Summary of proposed ($\mathbf{n}_\mathbf{f}, \mathbf{n}_\mathbf{c})$ phase structure of adjoint QCD:}} Having now filled the gap for the $n_f=3$ theory with a candidate phase we cannot help but briefly discuss some important characteristics of the emerging possible phase diagram in the $(n_c,n_f)$ plane (we ignore any possible issues regarding large $n_c$ in what follows). 

Consider keeping $n_c$ fixed and vary $n_f$. For $n_f>5$ the theory is IR free and all 0-form global symmetries are unbroken. The 1-form center symmetry is broken, due to the  perimeter law for the fundamental Wilson loop. As we lower $n_f$ to $n_f= 5$ and $n_f=4$ the theory is an interacting conformal field theory (CFT) in the IR, again with all 0-form (1-form) global symmetries unbroken (broken). Then at $n_f=3$ massless composite fermions are formed and discrete chiral symmetry is broken, while continuous chiral symmetry is still unbroken. The theory confines fundamental charges and the 1-form center symmetry is restored. In addition there is a TQFT, originating in the discrete symmetry breaking, to match the mixed $\mathbb{Z}_{2 n_fn_c}$-$\mathbb{Z}_{n_c}^{(1)}$ anomaly. Then at $n_f=2$ the likely \cite{Cordova:2018acb} (but see also the alternatives enumerated in \cite{Wan:2018djl}) scenario is that a fermion bilinear condensate is formed, breaking both continuous and discrete chiral symmetry. Finally, the $n_f=1$ theory is supersymmetric pure Yang-Mills theory with confinement and discrete chiral symmetry breaking.

Loosely speaking, we see that as we lower $n_f$ the theory prefers to break more and more of its global 0-form symmetries. This is somewhat reminiscent of the $(n_f, n_c)$ phase diagram of $\mathcal{N}=1$ supersymmetric QCD with fundamental chiral supermultiplets and may be, heuristically, what one expects. In Table \ref{IRphases} we provide a summary of the different IR phases.

{\flushleft\textbf{Acknowledgments:}} We thank Aleksey Cherman and Mikhail Shifman for discussions and suggestions. EP is supported by a Discovery Grant from NSERC. TAR is partially supported by the Danish National Research Foundation under the grant DNRF:90.

\bibliographystyle{apsrev4-1}

\begin{thebibliography}{99}

\bibitem{Anber:2018tcj} 
  M.~M.~Anber and E.~Poppitz,
  ``Two-flavor adjoint QCD,''
  Phys.\ Rev.\ D {\bf 98}, no. 3, 034026 (2018)
  [arXiv:1805.12290 [hep-th]].
  
  
\bibitem{Cordova:2018acb} 
  C.~Cordova and T.~T.~Dumitrescu,
  ``Candidate phases for SU(2) adjoint QCD$_4$ with two flavors from $\mathcal{N}=2$ supersymmetric Yang-Mills theory,''
  arXiv:1806.09592 [hep-th].
  
\bibitem{Bi:2018xvr} 
  Z.~Bi and T.~Senthil,
  ``An adventure in topological phase transitions in 3+1-D: non-abelian deconfined quantum criticalities and a possible duality,''
  arXiv:1808.07465 [cond-mat.str-el].
  
\bibitem{Wan:2018djl} 
  Z.~Wan and J.~Wang,
  ``Adjoint QCD$_4$, deconfined critical phenomena, symmetry-enriched topological quantum field theory, and higher symmetry extension,''
  Phys.\ Rev.\ D {\bf 99}, no. 6, 065013 (2019)
  [arXiv:1812.11955 [hep-th]].
  
  
\bibitem{Sannino:2004qp} 
  F.~Sannino and K.~Tuominen,
   ``Orientifold theory dynamics and symmetry breaking,''
  Phys.\ Rev.\ D {\bf 71}, 051901 (2005)
  [hep-ph/0405209].
  

\bibitem{Catterall:2007yx} 
  S.~Catterall and F.~Sannino,
  ``Minimal walking on the lattice,''
  Phys.\ Rev.\ D {\bf 76}, 034504 (2007)
  [arXiv:0705.1664 [hep-lat]].

  
  
  
\bibitem{Unsal:2007vu} 
  M.~Unsal,
  ``Abelian duality, confinement, and chiral symmetry breaking in QCD(adj),''
  Phys.\ Rev.\ Lett.\  {\bf 100}, 032005 (2008)
  [arXiv:0708.1772 [hep-th]].
  
\bibitem{Unsal:2007jx} 
  M.~Unsal,
  ``Magnetic bion condensation: a new mechanism of confinement and mass gap in four dimensions,''
  Phys.\ Rev.\ D {\bf 80}, 065001 (2009)
  [arXiv:0709.3269 [hep-th]].

\bibitem{Cherman:2018mya} 
  A.~Cherman, M.~Shifman and M.~Unsal,
  ``Bose-Fermi cancellations without supersymmetry,''
  arXiv:1812.04642 [hep-th].


  
\bibitem{Caswell:1974gg} 
  W.~E.~Caswell,
  ``Asymptotic behavior of nonabelian gauge theories to two loop order,''
  Phys.\ Rev.\ Lett.\  {\bf 33}, 244 (1974).


\bibitem{Banks:1981nn} 
  T.~Banks and A.~Zaks,
  ``On the phase structure of vector-like gauge theories with massless fermions,''
  Nucl.\ Phys.\ B {\bf 196}, 189 (1982).

 
\bibitem{DelDebbio:2010hx} 
  L.~Del Debbio, B.~Lucini, A.~Patella, C.~Pica and A.~Rago,
  ``The infrared dynamics of Minimal Walking Technicolor,''
  Phys.\ Rev.\ D {\bf 82}, 014510 (2010)
  [arXiv:1004.3206 [hep-lat]].
  
\bibitem{Hietanen:2009az} 
  A.~J.~Hietanen, K.~Rummukainen and K.~Tuominen,
  ``Evolution of the coupling constant in SU(2) lattice gauge theory with two adjoint fermions,''
  Phys.\ Rev.\ D {\bf 80}, 094504 (2009)
  [arXiv:0904.0864 [hep-lat]].

\bibitem{Bergner:2016hip} 
  G.~Bergner, P.~Giudice, G.~Munster, I.~Montvay and S.~Piemonte,
  ``Spectrum and mass anomalous dimension of SU(2) adjoint QCD with two Dirac flavors,''
  Phys.\ Rev.\ D {\bf 96}, no. 3, 034504 (2017)
  [arXiv:1610.01576 [hep-lat]].


\bibitem{Ryttov:2016hdp} 
  T.~A.~Ryttov,
  ``Consistent perturbative fixed point calculations in QCD and supersymmetric QCD,''
  Phys.\ Rev.\ Lett.\  {\bf 117}, no. 7, 071601 (2016)
  [arXiv:1604.00687 [hep-th]].

\bibitem{Ryttov:2017kmx} 
  T.~A.~Ryttov and R.~Shrock,
  ``Higher-order scheme-independent series expansions of $\gamma_{\bar\psi\psi,IR}$ and $\beta'_{IR}$ in conformal field theories,''
  Phys.\ Rev.\ D {\bf 95}, no. 10, 105004 (2017)
  [arXiv:1703.08558 [hep-th]].
  
\bibitem{Ryttov:2017lkz} 
  T.~A.~Ryttov and R.~Shrock,
  ``Physics of the non-Abelian Coulomb phase: insights from Pade approximants,''
  Phys.\ Rev.\ D {\bf 97}, no. 2, 025004 (2018)
  [arXiv:1710.06944 [hep-th]].
  
  
\bibitem{Bergner:2017gzw} 
  G.~Bergner, P.~Giudice, G.~Munster, P.~Scior, I.~Montvay and S.~Piemonte,
  ``Low energy properties of SU(2) gauge theory with N$_{f}$ = 3/2 flavours of adjoint fermions,''
  JHEP {\bf 1801}, 119 (2018)
  [arXiv:1712.04692 [hep-lat]].



\bibitem{Cachazo:2002ry} 
  F.~Cachazo, M.~R.~Douglas, N.~Seiberg and E.~Witten,
  ``Chiral rings and anomalies in supersymmetric gauge theory,''
  JHEP {\bf 0212}, 071 (2002)
  [hep-th/0211170].
     
     
\bibitem{Ramond:2010zz} 
  P.~Ramond,
 ``Group theory: a physicist's survey,''
  Cambridge, UK: Univ. Pr. (2010) 310 p
  
\bibitem{Csaki:1997aw} 
  C.~Csaki and H.~Murayama,
  ``Discrete anomaly matching,''
  Nucl.\ Phys.\ B {\bf 515}, 114 (1998)
  [hep-th/9710105].
  
\bibitem{Komargodski:2011vj} 
  Z.~Komargodski and A.~Schwimmer,
  ``On renormalization group flows in four dimensions,''
  JHEP {\bf 1112}, 099 (2011)
  [arXiv:1107.3987 [hep-th]].

  
\bibitem{Gaiotto:2014kfa} 
  D.~Gaiotto, A.~Kapustin, N.~Seiberg and B.~Willett,
  ``Generalized global symmetries,''
  JHEP {\bf 1502}, 172 (2015)
  [arXiv:1412.5148 [hep-th]].
  
\bibitem{Gaiotto:2017yup} 
  D.~Gaiotto, A.~Kapustin, Z.~Komargodski and N.~Seiberg,
  ``Theta, time reversal, and temperature,''
  JHEP {\bf 1705}, 091 (2017)
  [arXiv:1703.00501 [hep-th]].
  
\bibitem{tHooft:1979rtg} 
  G.~'t Hooft,
  ``A property of electric and magnetic flux in nonabelian gauge theories,''
  Nucl.\ Phys.\ B {\bf 153}, 141 (1979).


\bibitem{vanBaal:1982ag} 
  P.~van Baal,
  ``Some results for SU(N) gauge fields on the hypertorus,''
  Commun.\ Math.\ Phys.\  {\bf 85}, 529 (1982).
  
\bibitem{Anber:2018xek} 
  M.~M.~Anber and E.~Poppitz,
  ``Domain walls in high-$T SU(N)$ super Yang-Mills theory and QCD(adj),''
  arXiv:1811.10642 [hep-th].
  
\bibitem{Witten:1979kh} 
  E.~Witten,
  ``Baryons in the $1/n$ Expansion,''
  Nucl.\ Phys.\ B {\bf 160}, 57 (1979).
  
\bibitem{Seiberg:1994pq} 
  N.~Seiberg,
  ``Electric - magnetic duality in supersymmetric nonAbelian gauge theories,''
  Nucl.\ Phys.\ B {\bf 435}, 129 (1995)
  [hep-th/9411149].
  
\bibitem{Schmaltz:1998bg} 
  M.~Schmaltz,
  ``Duality of nonsupersymmetric large N gauge theories,''
  Phys.\ Rev.\ D {\bf 59}, 105018 (1999)
  [hep-th/9805218].
  

\bibitem{Armoni:2003gp} 
  A.~Armoni, M.~Shifman and G.~Veneziano,
  ``Exact results in non-supersymmetric large N orientifold field theories,''
  Nucl.\ Phys.\ B {\bf 667}, 170 (2003)
  [hep-th/0302163].

\bibitem{Armoni:2003fb} 
  A.~Armoni, M.~Shifman and G.~Veneziano,
  ``SUSY relics in one flavor QCD from a new 1/N expansion,''
  Phys.\ Rev.\ Lett.\  {\bf 91}, 191601 (2003)
  [hep-th/0307097].



\bibitem{Unsal:2006pj} 
  M.~Unsal and L.~G.~Yaffe,
  ``(In)validity of large N orientifold equivalence,''
  Phys.\ Rev.\ D {\bf 74}, 105019 (2006)
  [hep-th/0608180].
  
\bibitem{Armoni:2007kd} 
  A.~Armoni, M.~Shifman and M.~Unsal,
  ``Planar limit of orientifold field theories and emergent center symmetry,''
  Phys.\ Rev.\ D {\bf 77}, 045012 (2008)
  [arXiv:0712.0672 [hep-th]].

\bibitem{Preskill:1981sr} 
  J.~Preskill and S.~Weinberg,
  ``'Decoupling' constraints on massless composite particles,''
  Phys.\ Rev.\ D {\bf 24}, 1059 (1981).
\end{thebibliography}

\end{document}